\documentclass[10pt, letterpaper]{article} 

\usepackage {opex3} 
\usepackage {graphicx} 
\usepackage {graphicx} 
\usepackage {amssymb} 
\usepackage {amsmath}
\usepackage {ulem}
\usepackage{cite}
\newcommand{\reffig}[1]{Fig.~\ref{#1}}
\newcommand{\refeq}[1]{Eq.~(\ref{#1})}

\newcommand{\vect}[1]{\mathrm{\mathbf{#1}}} 
\newcommand{\pderiv}[2]{\frac{\partial#1}{\partial#2}} 

\begin{document}

\title{Generation of terahertz radiation from ionizing two-color laser
  pulses in Ar filled metallic hollow waveguides}

\author{I. Babushkin$^1$, S. Skupin$^{2,3}$, and J. Herrmann$^4$}
\address{$^1$ Weierstra\ss -Institut f\"ur Angewandte Analysis und
  Stochastik, 10117 Berlin, Germany \\ $^2$ Max Planck Institute for
  the Physics of Complex Systems, 01187 Dresden, Germany\\
  $^3$ Friedrich Schiller University, Institute of Condensed Matter
  Theory and Optics, 07742 Jena, Germany\\ $^4$ Max Born Institute for
  Nonlinear Optics and Short Time Spectroscopy, 12489, Berlin,
  Germany}

\date{\today}

\begin{abstract}
  The generation of THz radiation from ionizing two-color femtosecond
  pulses propagating in metallic hollow waveguides filled with Ar is
  numerically studied. We observe a strong reshaping of the
  low-frequency part of the spectrum.  Namely, after several
  millimeters of propagation the spectrum is extended from hundreds of
  GHz up to $\sim 150$~THz. For longer propagation distances, nearly
  single-cycle near-infrared pulses with wavelengths around 4.5~$\mu$m
  are obtained by appropriate spectral filtering, with an efficiency
  of up to 0.25~\%.
\end{abstract}





\section{Introduction}

In recent years, the range of wavelengths where coherent radiation can
be generated has grown dramatically into both high and low frequency
domain. Remarkably, most of the methods to obtain radiation at extreme
frequencies use, in one or the other way, nonlinear processes in
laser-induced plasma. One prominent example is high harmonic
generation (HHG) where frequencies thousand times larger than the
frequency of the pump pulse are excited, exploiting the recollision
dynamics of electrons ionized by the intense light pulses. More
recently it was demonstrated that a two-color fs beam allows
generation of new frequencies just in the opposite part of the
spectrum, namely in the THz range, hundreds times smaller than the
optical pump frequency. 
To this end, a short two-color pulse of fundamental frequency and
second harmonic is strongly focused into a plasma spot \cite{cook00,
  bartel05, xie06, dai06, reimann07, roskos07, kim07, kim08a, wang08,
  chen08, kim09, dai09} The observed THz emission generated in this
scheme has been attributed to the laser-induced plasma current in the
asymmetric two-color field \cite{kim07, kim08a}.
Using such scheme generation of strong THz radiation was reported,
with a spectrum which can be as broad as 70~THz \cite{kim08a}. Such
broad-band coherent radiation can allow new applications providing the
possibility to probe complex molecules or as an analytical and imaging
tool in a broad range of fields.

In this article we consider a modification of the above described
focusing geometry in a bulk gas by using a metallic hollow waveguide
with a cladding from aluminum to guide both THz and optical
radiation. The use of a waveguide prevents the diffraction of light
for all wavelengths involved.  In addition, such setup allows to
realize nearly single-mode operation for a wide range of frequencies.
High threshold intensities for ionization-induced processes in the
range larger than 100~TW/cm$^2$ have been already realized in the
context of HHG schemes in hollow waveguides \cite{durfee99, gibson03}.
We will show that during propagation a dramatic spectral broadening of
the low-frequency part of the spectrum occurs, caused by only modest
changes in the fundamental and second harmonic fields.  In particular,
ultrashort pulses of 10~fs duration lead to the formation of
low-frequency spectrum extended from several hundreds of GHz to
approximately 150~THz. Upon further propagation we report the
generation of nearly single-cycle pulses with central wavelengths
around 4.5~$\mu$m and efficiency of the order of 0.25~\%.


\section{The model}

In the previous works, THz emission by two-color fields in gases has
been interpreted by four-wave mixing rectification
\cite{cook00,bartel05,xie06,dai06,reimann07}. In contrast, later
studies attributed THz emission to the laser-induced plasma-current in
the asymmetric two-color field
\cite{kim07,kim08a,wang08,chen08,kim09,dai09} described by the local
quasi-classical model of the electron current in the laser field
without taking into account propagation effects.  As an alternative
approach, particle-in-cell simulations were used
\cite{wang08,chen08}. Those methods are, however, computationally very
expensive, and propagation effects have been taken into account over
distances of a few micrometers only.
In such models the microscopic current is described as a sum over
contributions from all electrons, born at discrete times $t_n$. The
''continuous'' analog of this approach is given by
(\cite{kim07,kim08a,kim09})
\begin{equation}
  \vect  J(t) =
  q\int\limits_{-\infty}^t \vect  v(t,t_0)
  \dot{\rho}(t_0) dt_0,\label{eq:1}
\end{equation}
where $q$ is the electron charge. 
The current density $\vect J(t)$ at the time $t$ is obtained by
integrating over all impacts from different ionization times $t_0$:
$\vect v(t,t_0)$ is the velocity of electrons at time $t$ which were
born at time $t_0$, and $\delta \rho(t_0) = \rho(t_0+\delta t) -
\rho(t_0) \approx \dot{\rho}(t_0) \delta t$ is the density of
electrons born at time $t_0$ \footnote{We use the convention $\dot{f}
  \equiv \pderiv{f}{t}$.}.  This approach was utilized in the previous
works of Kim and coauthors \cite{kim07,kim08a,kim09}.

The additional integration over the variable $t_0$ increases the
numerical effort, but one can eliminate the microscopic velocity
distribution $\vect v(t,t_0)$ and derive a differential equation for
the macroscopic current density $\vect J(t)$.  Assuming no interaction
with other particles and zero velocity of new-born electrons, the
equation for the electron velocity can be written as $\vect v(t,t_0) =
\frac{q}{m_e} \int_{t_0}^t \vect E(\tau) d\tau$ (where $\vect E$ is
the electric field, $m_e$ is the electron mass).  With the simple
additive property of integrals we can rewrite it as $ \vect v(t,t_0) =
\vect v(t,-\infty)-\vect v(t_0,-\infty)$.  Substituting this into
\refeq{eq:1} and differentiating with respect to $t$ we obtain $
\dot{\vect J}(t) = q \dot{\vect v}(t) \rho(t)$.  It is easy to see
that by introducing new variables $\vect v' = \vect v \exp(-\nu_e t)$,
$\vect J'=\vect J \exp(-\nu_e t)$ (where $\nu_e$ is the electron-ion
collision rate) and proceeding analogously we can include collisional
effects as well, and finally obtain
\begin{equation}
  \label{eq:14}
  \dot{\vect J}(t) + \nu_e \vect J(t) = \frac{q^2}{m_e} \vect E(t)
  \rho(t).
\end{equation}
Indeed, \refeq{eq:14} coincides with the well known (and very commonly
used) equation for the current in plasmas \cite{berge07,sprangle04}.

In the present article, the static model for the tunneling ionization
\cite{landau65} is used:

  \begin{equation}
  \label{plasma}
  \dot{\rho}(t) = W_{\rm ST}[E(t)] [\rho_{\rm at} - \rho(t)], \quad \quad
  W_{ST}[E(t)] = \alpha \frac{E_a}{E(t)}
  \exp\left\{-\beta \frac{E_a}{E(t)}\right\}.
  \end{equation}
  Here, $\rho_{\rm at}$ is the neutral atomic density, and $E_a
  \approx 5.14 \times 10^{11}$~Vm$^{-1}$ the atomic field; the
  coefficients $\alpha = 4 \omega_a r_H^{5/2}$ and $\beta = (2/3)
  r_H^{3/2}$ are defined through the ratio of ionization potentials of
  argon and hydrogen atoms $r_H=U_{Ar}/U_H$, $U_{Ar}=15.6$~eV,
  $U_H=13.6$~eV, and $\omega_a \approx 4.13 \times 10^{16}$ s$^{-1}$
  \cite{roskos07,landau65}.

  To describe the light propagation, we use the so called forward
  Maxwell equation (FME) for the fast oscillating optical field $\vect
  E(\vect r, z, t)$ in the
  waveguide with the optical axis along $z$ direction \cite{husakou01}:
%
\begin{equation}
  \pderiv{\tilde{ \vect E} (\vect r,z,\omega)}{z} =   i k(\omega) 
  \tilde{\vect E} (\vect
  r,z,\omega) + \frac{i}{2 k(\omega)} \Delta_\bot \tilde{\vect E}
  (\vect r,z,\omega) + \frac{i \mu_0 \omega c}{2 n(\omega)} \tilde{\vect P}_{nl}
  (\vect r,z,\omega), 
  \label{eq:fme-vect}
\end{equation}
where $\vect r = \{x,y\}$, $\Delta_\bot = \partial_{xx} + \partial_{yy}$, $n(\omega)$ is
the refractive index of Ar, $k(\omega)=n(\omega) \omega/c$. 
$\tilde{\vect E}(\vect r,z, \omega)$ and $\tilde{\vect P_{nl}}(\vect
r,z, \omega)$ are the Fourier transforms of $\vect E(\vect r,z,
t)$, $\vect P_{nl}(\vect r,z, t)$ with respect to time,
$\vect P_{nl}$ is
the nonlinear polarization, which includes both the Kerr ($\vect
P_{Kerr}=\varepsilon_0 \chi^{(3)}|\vect E|^2 \vect E$) nonlinearity
and the plasma contribution: $\tilde{\vect P}_{nl} = \tilde{\vect
  P}_{Kerr} + i\tilde{ \vect J}/\omega + i\tilde {\vect
  J}_{loss}/\omega$, where $\vect J_{loss} =W_{\mathrm{ST}}(\rho
_{\mathrm{at}}-\rho)U_{Ar}/\vect E$ is a loss term accounting for
photon absorption during ionization; ``division by a vector'' means
the component-wise division \cite{landau65}.  
Decomposing the field
into a series of linear eigenmodes of the waveguide $\{\vect F_j(\vect
r), \, j=1,\: \ldots \: \infty\}$ with corresponding propagation
constants $\beta_j(\omega)$, we obtain the following set of equations
for the amplitudes of the eigenmodes $E_j$, coupled through the
nonlinear polarization term $P_{nl}^{(j)} = \int \vect F_j \vect
P_{nl} d \vect r$:
  \begin{equation}
    \label{eq:fme-base}
    \pderiv{\tilde{E}_j(z,\omega)}{z} = i \beta_j(\omega)  
    \tilde{E}_j(z,\omega) + i \frac{c \mu_0 \omega^2}{2
      k(\omega)} \tilde{P}^{(j)}_{nl}(z,\omega).
  \end{equation}

\section{Numerical simulations}

In our simulations, we assume a dielectric waveguide with aluminum
coating of the inner walls and diameter $d=100$~$\mu$m. Such
waveguides can be routinely produced (see, e.g., \cite{tzankov07}).
The main purpose of the waveguide is not to modify the dispersion
relation but to confine both pump and THz radiation (especially the
latter, because it quickly spreads out otherwise as a result of
diffraction \cite{subm10}).  Taking into account that THz as well as
fundamental and second harmonic frequencies undergo strong absorption
due to plasma generation a long waveguide is not useful. Therefore, we
restrict our simulations to distances $\le 1$~cm.

\begin{figure}
\begin{center}
\includegraphics[width=0.8\columnwidth]{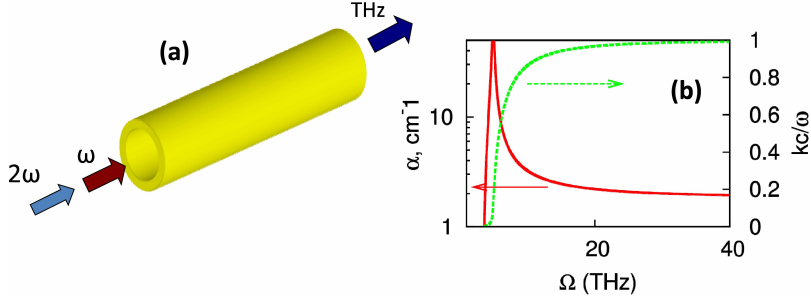}
\caption{ \label{fig:1} (Color online). (a) Scheme for THz generation
  considered in this article.  A short pulse with center frequency
  $\nu_0= 375$~THz (corresponding to wavelength $\lambda=800$~nm) and
  its second harmonic are focused into an aluminum coated hollow
  waveguide.  The THz emission is detected at the exit of the
  waveguide. (b) Losses $\alpha$ (red solid line) and refractive
  index change $\delta n = kc/\omega$ (green dashed line)
  induced by the Al waveguide with the diameter $d=100$~$\mu$m for the
  mode EH$_{11}$ versus frequency $\Omega=\omega/2\pi$. }
\end{center}
\end{figure}

The linearly-polarized waveguide mode EH$_{11}$ has no cut-off at low
frequencies, and the fundamental and second-harmonic frequencies can
be coupled into the waveguide with almost 100~\% efficiency from the
free-space propagating laser mode.  The dispersion and losses for the
mode EH$_{11}$ of the 100-$\mu$m waveguide are shown in
\reffig{fig:1}(b) in low-frequency range. Both quantities are
calculated using the direct solution of the corresponding boundary
value problem \cite{markatili64}, Drude model for the dispersion in
aluminum \cite{bass94} as well as Sellmeier formula for the dispersion
in argon \cite{leonard74}.  One can see that the losses are relatively
large only in the small frequency range corresponding to wavelengths
of the order of the waveguide diameter.  The aluminum coating was
chosen because then the modeling of linear dispersion and losses in
the whole frequency range from sub-THz to sub-PHz is straight forward
using the Drude model \cite{bass94}. In contrast, for dielectric
coatings (such as fused silica) such simple description for the whole
spectral region of interest fails. Nevertheless, we expect even weaker
losses than for the metallic waveguide.

\begin{figure*}
\begin{center}
\includegraphics[width=1\textwidth]{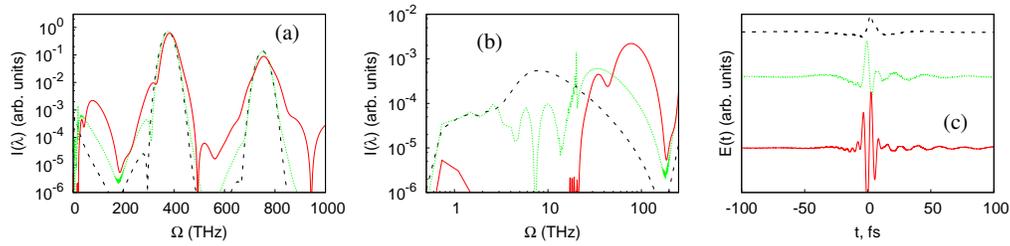}
\caption{ \label{fig:2} (Color online). Evolution of the THz spectrum upon propagation
  for a waveguide with diameter $d=100$ $\mu$m, input intensities
  $I_\omega=10^{14}$~W/cm$^2$, $I_{2\omega}=2\times 10^{13}$~W/cm$^2$
  and pulse duration 10~fs. (a) The spectrum and (b) low-frequency
  part of spectrum (note the logarithmic scale of y-axis in (a) and
  both axes in (b)) for $z=0.5$, 2, and 10~mm of propagation (black
  dashed, green dotted and red solid lines). (c) Pulses obtained by
  filtering the long-wavelength part of the spectrum below 250
  THz. Propagation distances, color coding and line styles are the
  same as in (a).  }
\end{center}
 \end{figure*}

Results from simulations of ultrashort pulses with 10~fs duration and
frequencies $\nu_0=375$ THz (corresponding to the wavelength
$\lambda=800$~nm) and $2\nu_0$ , having intensities
$I_\omega=10^{14}$~W/cm$^2$ and $I_{2\omega}=2\times10^{13}$~W/cm$^2$
correspondingly are shown in \reffig{fig:2}. We checked that the
transverse field distribution consists predominantly of the
fundamental and few higher order modes.
 At the initial stage of propagation, broadband THz radiation is
 generated from approximately 0.5~THz to 80~THz (see \reffig{fig:2}(a)
 and (b)). With further propagation (up to 2~mm) the low-frequency
 part of the spectrum grows in the direction to higher frequencies (up
 to 150~THz).  At the same time, the spectral intensity around 10~THz
 degrades because of the strong losses of the waveguide mode. From
 2~mm up to $\sim 7$~mm the long-wavelength part of the spectrum
 shifts continuously towards higher frequencies. Finally, starting
 from 7~mm the spectrum does not change considerably anymore, except
 for its decrease due to losses. At this propagation distance, the
 long-wavelength part of the spectrum is localized around 60-80 THz,
 with the center around 66 THz (corresponding to a wavelength $\sim
 4.5$~$\mu$m). The observed efficiency of this ''frequency
 down-conversion'' reaches 0.25 percent.

 The electric field obtained by filtering the long-wavelength part of
 the spectrum for different propagation distances is shown in
 \reffig{fig:2}(c). One can see that despite the strong reshaping of
 the spectrum, the envelope of the obtained THz pulses does not change
 significantly.  The amplitude of the pulse grows considerably during
 propagation, from approximately 1~\% of the pump amplitude at
 $z=0.5$~mm (which correspond to the conversion efficiency $\sim
 10^{-4}$) to about 5~\% at $z=10$~mm (which correspond to the
 conversion efficiency $\sim 2.5\times 10^{-3}$).

 \reffig{fig:2}(a) reveals that in contrast to the strong reshaping of
 the low frequency components, the spectrum in the range of the pump
 frequencies remains almost unchanged.  Nevertheless, the spectrum of
 the pump broadens slightly upon propagation. Due to plasma-induced
 refraction index change the maximum frequency of the fundamental
 pulse is slightly shifted to the blue \cite{wood91, rae92}. In our
 case the shift is small, only about 10 THz. As elaborated in
 \cite{subm10}, even small changes in the pump fields can affect
 dramatically the spectral shape of the generated low-frequency
 radiation.  Such changes affect significantly the temporal positions
 of the ionization events.  Contributions from the electrons born in
 different ionization events are superimposed, and interference
 effects lead to a reshaping the THz spectra.

\begin{figure}[htbp!]
\begin{center}
\includegraphics[width=0.9\columnwidth]{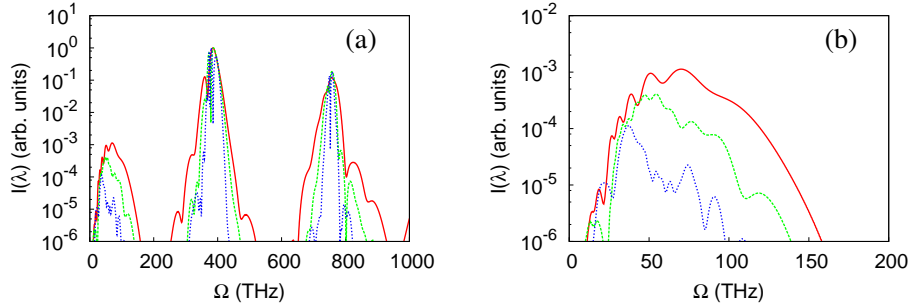}
\caption{ \label{fig:dd} (Color online). Spectra for different input
  pulse durations (25, 50 and 1000~fs are shown by red solid, green
  dashed and blue dotted lines, respectively) for 1 cm of
  propagation. In (a) and (b) the full spectra and their low-frequency
  parts are shown, respectively. Spectra are normalized to their
  maximum value, so that efficiencies can be compared.  }
\end{center}
 \end{figure}

 The situation is qualitatively similar for other pulse durations (see
 \reffig{fig:dd}). In general, the efficiency of the THz generation is
 larger for shorter pulse durations (in \reffig{fig:dd}, spectra are
 normalized to unity at their maxima, so the efficiencies can be
 directly compared). This observation can be explained by the
 additional asymmetry in the current which is introduced by the finite
 pulse duration and therefore more pronounced for short pulses.

\section{Conclusion}

In conclusion, we studied THz generation by plasma currents through
noble gas ionization in strong asymmetric laser fields. We have shown
that the mechanism allows the relatively simple generation of
extremely wide range of frequencies, starting from THz to near
infrared. By using hollow waveguides we overcome the strong
diffraction of the generated low-frequency generation.  We report that
depending on the length of the waveguide either THz supercontinuum or
almost single-cycle near-infrared pulses can be produced at output
with $\sim 0.25$~\% efficiency.  We believe that our findings may open
new intriguing possibilities to control the spectral and temporal
shape of low-frequency radiation generated by the ionization dynamics
in intense optical fields.

 
\end{document}